\documentstyle [pre,aps,epsfig]{revtex}
\newenvironment{mytitle}{\begin{center} \LARGE}{\\ [.1in]\end{center}} 
\newenvironment{myauthor}{\begin{center} }{\\ [.1in]\end{center}} 
\newenvironment{myinstit}{\begin{center} \it}{\end{center}}

\begin{document}

\begin{mytitle}
Some Thermal and Electrical Properties of Candelilla Wax
\end{mytitle}
\vspace{.5cm}

\begin{myauthor}
V. Dossetti-Romero, J. A. M\'endez-Berm\'udez, and E. L\'opez-Cruz
\end{myauthor}

\begin{myinstit}
Instituto de F\'{\i}sica, Universidad Aut\'onoma de Puebla, Apartado Postal 
J-48, Puebla 72570, M\'exico
\end{myinstit}

\begin{abstract}
We report the values of some thermal and electrical properties of Candelilla Wax (euphorbia cerifera). The open-cell photoacoustic technique and another photothermic technique - based on the measure of the temperature decay of a heated sample - were employed to obtain the thermal diffusivity ($\alpha_{s} = 0.026 \pm 0.00095 \, \mbox{cm}^{2}\mbox{/sec}$) as well as the thermal conductivity ($k=2.132 \pm 0.16 \, \mbox{W/mK}$) of this wax. The Kelvin null method was used to measure the dark decay of the surface potential of the sample after a Corona Discharge, obtaining a resistivity of $\rho_e=5.98 \pm 0.19 \times 10^{17} \, \mbox{ohm-cm}$.
\end{abstract}

\vspace{.5cm}

\section{Introduction}
In past years a growing interest in the electronic and optical properties of organic materials has been shown \cite{1}. Some of the works have been centered in some particular aspects of the material under study \cite{2} as well as in certain applications \cite{3}. At the very beginning of the insulator and electret research one of the phenomena extensively studied was the Costa-Ribeiro effect \cite{4}, among the materials in which one can find some waxes (natural as well as synthetic ones) and in organic semiconductors \cite{5}.

In this work we are interested in studying some electrical and thermal properties of Candelilla wax, a natural wax from a bush wildly grown in northern Mexico. We measured the thermal diffusivity, conductivity, and capacity of this material using the standard method of photoacoustic spectroscopy \cite{6} combined with the measurement of thermal decay of a cooling process in vacuum \cite{Hatta} to obtain the value of the product: mass density and thermal capacity. The electric conductivity is studied employing the Corona Discharge and the Kelvin null method \cite{7}.

\section{Experimental Details}
The samples were small pieces of candelilla wax fused and cooled in order to shape them as platelets of $\sim 500$ microns in thickness and with an area of $\sim 1cm^2$. The measured fusion temperature of this wax is $T_c\simeq 69^\circ \mbox{C}$, which is in good agreement with the $67-69^\circ \mbox{C}$ reported in the literature \cite{11}.

\subsection{Open-cell photoacoustic technique}
The photoacoustic technique used in this work was the open cell method widely reported in the literature \cite{8}. The thermal diffusivity $\alpha_s$, was measured using the experimental arrangement shown in Fig.\ 1. The sample is directly mounted onto a commercial electret microphone. The beam of a 170 W tungsten lamp was focused onto the sample and mechanically chopped. As a result of the periodic heating of the sample by the absorption of the modulated light, the microphone produces a signal that was monitored with a lock-in amplifier as a function of modulation frequency. The temperature reached by the periodically illuminated sample was $44^\circ \mbox{C}$, which is far below the fusion temperature $T_c$. 

\vspace{.5cm}
\begin{figure}[htb]
\begin{center}
\epsfig{file=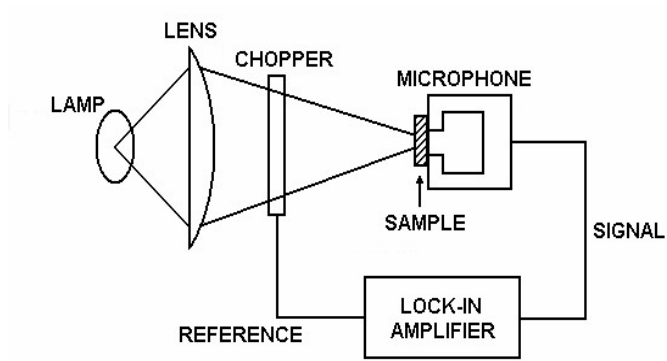,width=3.5in,height=2in}
\caption{Experimental arrangement used to measure the thermal diffusivity $\alpha_{s}$.}
\end{center}
\end{figure}

\subsection{Description of the method used for measuring $\rho c$}
Fig.\ 2 shows the experimental set-up used for measuring the product of the mass density and the specific heat \cite{Hatta}. Prior to the measurements both faces of the sample are sprayed with black paint in order to make its emissivity approximately equal to one. The sample is positioned inside the vacuum chamber with one of its faces (which we will call front face) illuminated by the light beam of a 60 W tungsten lamp properly focused. The temperature of the back face (the non-illuminated face) of the sample is traced with a Cu-constantan type thermopar connected to a temperature monitor while the temperature increases up to its equilibrium value, about 26 degrees above room temperature. Later on, the light is interrupted and the temperature is traced out, acquiring its value while the sample cools down up to its equilibrium value at room temperature.

\vspace{.5cm}
\begin{figure}[htb]
\begin{center}
\epsfig{file=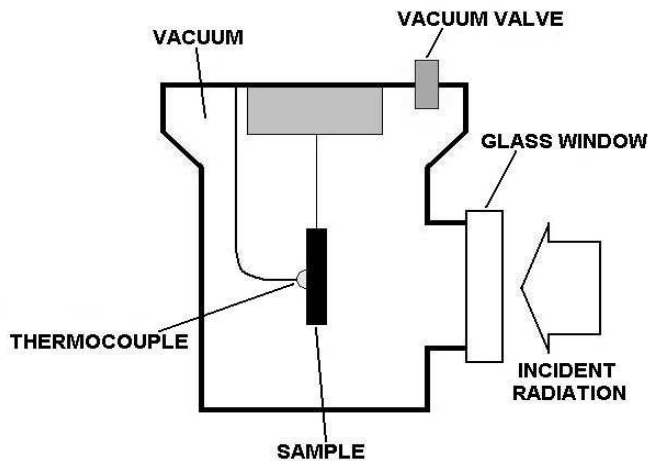,width=3.5in,height=2.5in}
\caption{Experimental arrangement used to measure the product $\rho c$.}
\end{center}
\end{figure}

\subsection{Corona dicharge and Kelvin null method} 
The experimental set up for measuring the resistivity is a standard one \cite{7} as seen on Fig.\ 3. The sample was provided with an ohmic contact on the back face and mounted with the free surface upward, as shown in the figure. It was placed on an arm that could be moved on the horizontal plane. In one position (Fig.\ 3a) it was charged by a negative 30 kV corona discharge in air, and in the other (Fig.\ 3b) the surface potential was measured by a Kelvin method. The discharge was driven from the tip of a fine wire using a DC voltage amplifier. The tip was positioned about 5 mm above the sample. After charging the sample, this was moved into the measuring position under the tracing electrode. The electrode consists of a conducting metal plate of about the same size of the sample and was driven back and forth in a vertical motion by a mechanical setup at a rate of approximately 1 Hz over a path of 1 mm. The sample was positioned in such a way as to be 1 mm away from the tracing electrode plate at the closest approach. This vibrating capacitor gave an output signal that was detected using an electrometer. Then the signal was balanced to a null voltage when the two plates (electrode and sample) are at the same potential. The balancing voltage was driven by a variable high DC voltage power supply. All the system was shielded by a Faraday box.

\vspace{.5cm}
\begin{figure}[htb]
\begin{center}
\epsfig{file=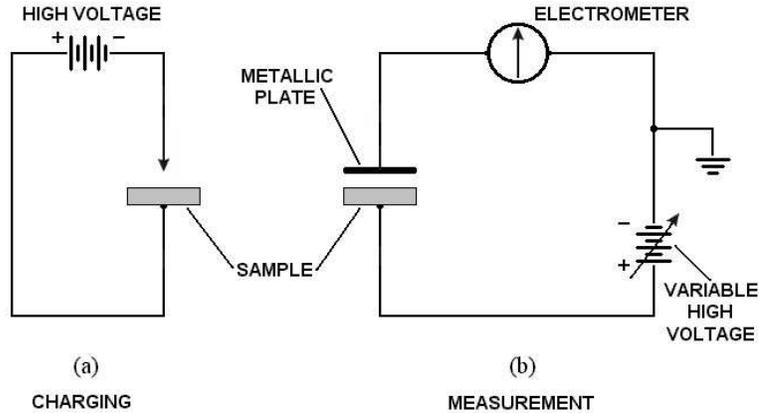,width=4in,height=2.2in}
\caption{Experimental arrangement used to measure the electrical resistivity $\rho_e$.}
\end{center}
\end{figure}

\section{Results and Discussion}

\subsection{Measurement of the thermal diffusivity}
In the open-cell photoacoustic technique, it is well known that the acoustic signal has two main contributions, one coming from the thermal diffusion phenomenon and the other one from the thermoelastic bending effect \cite{Rosencwaig,new,Perondi}. In orther to differenciate which one of these two contributions dominates in generating the photoacoustic (PA) signal, one has to compare the experimental measurement with the expression that describes those contributions. By means of this comparison we have found that the thermoelastic bending effect is predominant in the generation of the PA signal when it comes to samples of candelilla wax. Once identified where the main contribution to the PA signal comes from, one can calculate the thermal diffusivity $\alpha_s$ from the modulation frequency dependance of the signal phase. For a thermally thick sample, the expression for the pressure fluctuations inside the PA chamber induced by the thermoelastic bending effect is
\begin{equation}
p_{el} \simeq \frac{3 \alpha_{T} R^{4} \gamma P_{0} I_{0} \alpha_{s}}{4 \pi R_{c}^{2} I_{s}^{2} l_{g} k_{s} f}
\left[ \left( 1 - \frac{1}{x} \right)^{2} + \frac{1}{x^{2}} \right]^{1/2} e^{\, j \left[ \omega t \, + \,
\left( \pi/2 \right) \, + \, \phi \right]}
\label{eq:pressure}
\end{equation}
where $\alpha_{T}$ is the linear thermal expansion coefficient, $R$ is the microphone inlet 
hole radius, $\gamma$ is the air specific heat ratio, $P_{0}$ is the ambient pressure, $I_{0}$ is 
the absorbed light intensity, $R_{c}$ is the radius of the PA chamber in front of the diaphragm, 
$f$ is the modulation frequency, $x = l_{s}(\pi f/\alpha_{s})^{1/2}$, $l_{i}$, $k_{i}$ and $\alpha_{i}$ are the length, thermal conductivity, and the thermal diffusivity of material $i$, with subscripts $g$ and $s$ standing for gas media and sample respectively, and $\tan \phi = 1/(x-1)$.

From equation (\ref{eq:pressure}) one gets that the thermoelastic contribution to the PA signal amplitude, at high modulation frequencies ($x \gg 1$), varies as $f^{-1}$ and its phase $\phi_{el}$ approaches $90^{o}$ as
\begin{equation}
\phi_{el} \simeq \pi/2 + \arctan[1/(\sqrt{b_s f}-1)],
\label{eq:phase}
\end{equation}
here, $b_s$ is a fitting parameter. The other condition that must be fulfilled for an optically opaque sample to be thermally thick 
is $f \gg f_{c}$, where the cutoff frequency $f_{c}$ is given by $f_{c} = \alpha_{s} / (\pi l_{s}^{2})$.\

In a process where the main contribution to the PA signal comes from the thermal diffusion phenomenon, the almplitude and phase of the signal have a dependency on the frequency of the form $(1/f)\exp{-a_s\sqrt{f}}$ and $\phi_{th}=-(\pi/2)-a_s\sqrt{f}$, respectively, see Ref. \cite{Perondi}. Where $a_s$ is a fitting parameter which is related to the diffusivity $\alpha_s$ by $a_s = l_s\sqrt{\pi/\alpha_s}$.\

Figure 4 shows the amplitude and the phase of the PA signal for the candelilla wax sample. We can notice a good correspondence to the models presented before corresponding to the thermoelastic bending effect. In the case of the amplitude of the PA signal (Fig.\ 4a), it reproduces very well the $f^{-1}$ dependency with an exponent equal to $-1.0866$. Equation (\ref{eq:phase}) was used to fit the results shown in Fig.\ 4b (continuous line) together with our measured values (clear circles) for the phase of the PA signal. It is possible to estimate the thermal diffusivity $\alpha_{s}$ from the fitting parameter $b_s$ considering the relationship 
\begin{equation}
\alpha_{s} = \frac{\pi l_s^2}{b_s},
\label{eq:alpha}
\end{equation}
and using $l_{s}=643 \, \mu \mbox{m}$. In this case $f_{c} \simeq 2 \mbox{Hz}$, then from Fig.\ 4 we can see that we are in the thermally-thick-sample regime.

Finally we obtained the value $\alpha_{s} = 0.026 \pm 0.00095 \, \mbox{cm}^{2}\mbox{/sec}$ for the thermal diffusivity of candelilla wax.

\vspace{.5cm}
\begin{figure}[htb]
\begin{center}
\epsfig{file=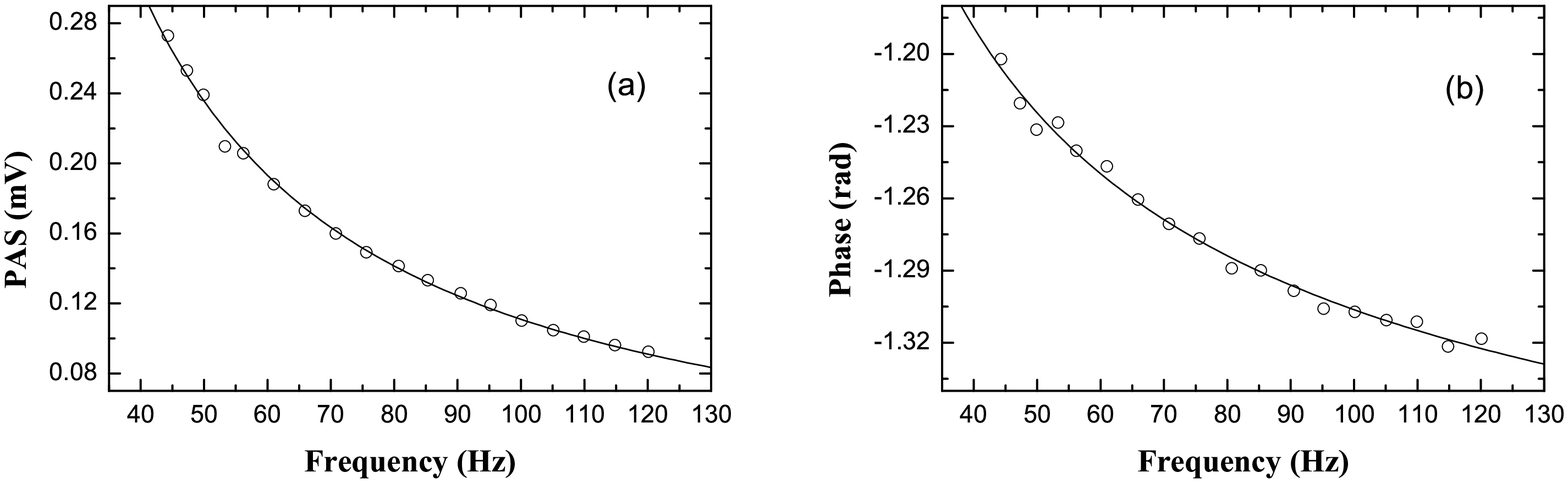,width=6.2in,height=2.7in}
\caption{Fitting of (a) the amplitude and (b) the phase of the photoacoustic signal as a function of the modulation frequency for a sample of candelilla wax.}
\end{center}
\end{figure}

\subsection{Measurement of $\rho c$ and computation of the thermal conductivity}
When one of the faces of the sample is illuminated as shown in Fig.\ 2, with a constant flux of light a lack of the equilibrium between the frontal (illuminated) and back (traced) faces of the sample is established. For the case where the width of the sample (including the two coats of black paint) $l$, is smaller than its transversal dimension, which is our case, this phenomenon can be described by a 1D equation. Thus, the conservation condition for the energy is

\begin{equation}
J_0 - \sigma T_1^4 - \sigma T_2^4 = \frac{d}{dt} \int_0^1 \rho c T(x,t) dx,
\label{eq3}
\end{equation}

\noindent where $J_0$ is the flux of incident light over the frontal face, $\sigma$ is the Stefan-Boltzmann constant, $T_1$ is the temperature of the frontal face, $T_2$ is the temperature of the back face, $\rho$ is the mass density of the sample and $c$ its specific heat at a constant pressure. In this equation we use explicitly the fact that the sample is painted with a thin coat of black paint that has an emissivity coefficient approximately equal to one \cite{Leon}.

\vspace{.5cm}
\begin{figure}[htb]
\begin{center}
\epsfig{file=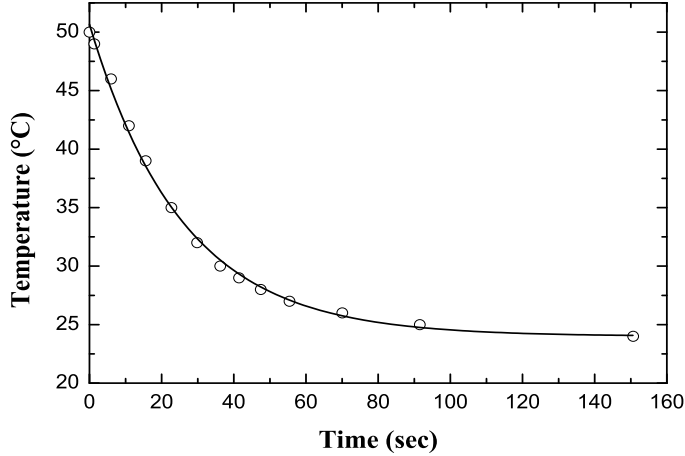,width=4in,height=2.8in}
\caption{Typical behaviour of the thermal decay of a candelilla wax sample as a function of time.}
\end{center}
\end{figure}

We define $\Delta T_i(t) = T_{i,max} - T_i(t)$, $(i=1,2)$, where $T_{1,max}$ and $T_{2,max}$ are the maximum temperatures reached by the frontal and back faces of the sample respectively, for long times when the equilibrium is reached. Substituting $\Delta T_i(t)$ in equation (\ref{eq3}) and linearizing the resultant equation in terms of $\Delta T_i/T_i$ we obtain

\begin{equation}
J_0 - \sigma T_{1,max}^4 - \sigma T_{2,max}^4 + 4\sigma T_{1,max}^3\Delta T_1(t) + 4\sigma T_{2,max}^3\Delta T_2(t) = \frac{d}{dt} \int_0^1 \rho c T(x,t) dx.
\label{linear}
\end{equation}

The sum of the first three terms of this equation is equal to zero, since for long times the flux of incident radiation and the flux of emitted radiation cancel out each other. The integral on the right hand side can be written as

\begin{equation}
\frac{d}{dt} \int_0^1 \rho c T(x,t) dx \approx \frac{\rho c l}{2} \frac{d}{dt} \left[ T_1(t)+T_2(t) \right] = -\frac{\rho c l}{2} \frac{d}{dt} \left[ \Delta T_1(t)+\Delta T_2(t) \right]  
\label{integ}
\end{equation}

\noindent using the fact that $c$ does not depend on the position and that it is practically constant in the interval of a few degrees above the room temperature. It is also a fact that for the values of $l$ and $J_0$ that we used in the laboratory it is fulfilled that $l \frac{dT(x,t)}{dt} \ll T_1(t) \cong T_2(t)$. From this condition we can assume that $\Delta T_1(t) \cong \Delta T_2(t)$, then for a decrease of the temperature from $T_{2,max}$ to $T_{2,0}$ after the light is interrupted, equation (\ref{eq3}) can be written as

\begin{equation}
8 \sigma T_{2,0}^3 \Delta T_2(t) = -\rho c l \frac{d\Delta T_2(t)}{dt}.
\end{equation}

Substituting the definition for $\Delta T_2(t)$ and using the boundary conditions $\Delta T_2(0) = 0$ and $\Delta T_2(\infty)=T_{2,max}-T_{2,0}$, we obtain the solution

\begin{equation}
T_2(t) = T_{2,0} + \left( T_{2,max}-T_{2,0} \right) \exp(-t/\tau_d)
\label{solution}
\end{equation}

\noindent for the decay of the temperature immediately after the illumination of the sample is interrupted. In this case the relaxation mean time $\tau_d$ is given by

\begin{equation}
\tau_d = \frac{\rho c l}{8 \sigma T_{2,0}^3}.
\label{rmt}
\end{equation}

Figure 5 presents the evolution of the temperature of the sample as a function of time in a typical thermal decay experiment. From the time constant $\tau_d = 26 \pm 1 \, \mbox{s}$, which fits very well to the relationship given by equation (\ref{solution}), one obtains the value of the product $\rho c$ from equation (\ref{rmt}). In our case $\rho c = 820030.44 \, \mbox{J/m}^3\mbox{K}$ with an error of $\pm 3.89$ percent, where $T_{2,0} = 23.88^\circ \mbox{C}$ and $l=377 \mu \mbox{m}$.

In order to obtain the thermal conductivity $k$, we can use the very well known relationship

\begin{equation}
\alpha_s = \frac{k}{\rho c},
\label{relation}
\end{equation}
which yields to $k=2.132 \pm 0.16 \, \mbox{W/mK}$.

\subsection{Measurement of electrical resistivity}
We consider a laminar sample with parallel surfaces (of thickness $l$ and surface area $A$) which after corona-charged behaves as an $RC$ circuit with time decay constant $\tau_e = RC$.

The resistance $R$ and the capacitance $C$ of the sample are given by $R=\rho_e(d/A)$ and $C=\xi(A/4\pi d) \times (1.1\times 10^{12})$, where $\rho_e$ is the electrical resistivity and $\xi$ is the dielectric constant ($R$ and $C$ are given in ohms and farads, respectively. $A$ and $d$ in centimeters). Then we have that $\tau_e = (\rho_e \xi /a\pi) \times (1.1\times 10^{12}) \, \mbox{s}$. We can make the approximation $\xi \approx  \pi$, valid for many materials \cite{10}, and obtain

\begin{equation}
\rho_e = \tau_e \times 10^{12} \, \mbox{ohm-cm}.
\label{density}
\end{equation}

It is known that the process for the discharge of a $RC$ circuit as a function of time is an exponential decay for the charge $q(t) = q_0 \exp(-t/\tau_e)$.\

A typical experiment of the dark decay of surface potential in a negatively charged candelilla wax sample is shown in Fig.\ 6, which is a plot of the surface potential as a function of time. Since the surface potential is proportional to the charge ($V=qC$), the experimental results in Fig.\ 6 can be fitted to the relationship

\begin{equation}
V(t) = V_0 \exp(-t/\tau_e).
\label{sp}
\end{equation}

The time constant obtained from these results was $\tau_e=166.17 \pm 5.48 \, \mbox{hours}$, and using equation (\ref{density}) one obtains a resistivity of $\rho_e=5.98 \pm 0.19 \times 10^{17} \, \mbox{ohm-cm}$.

\vspace{.5cm}
\begin{figure}[htb]
\begin{center}
\epsfig{file=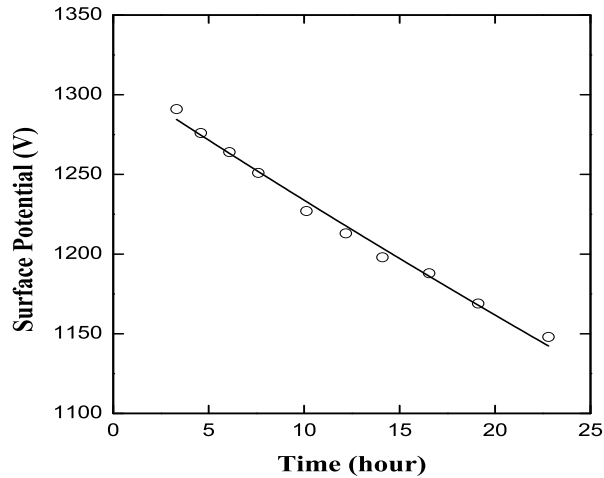,width=3.5in,height=2.8in}
\caption{Dark decay of the log of the surface potential as a function of time for a candelilla wax sample.}
\end{center}
\end{figure}

\section{Conclusions}
Although candelilla wax has a huge amount of industrial applications, some of its electric and thermal properties are not well understood. In some handbooks one can find the value of its dielectric constant but not its thermal diffusivity or its thermal conductivity, nor its resistivity \cite{11}. From the standard open-cell photoacoustic and thermal decay techniques one very easily finds some of the above-mentioned physical properties. Using equations (\ref{eq:alpha}) and (\ref{rmt}) we obtained the thermal diffusivity $\alpha_s$, and the product $\rho c$ respectively, starting with the values obtained from the fittings of the data to equations (\ref{eq:phase}) and (\ref{solution}). One can calculate the heat capacity once one measures the mass density $\rho$ of the sample. In our case the heat capacity was measured to be $c=754.71 \pm 29.35 \, \mbox{J/kgK}$ and the thermal diffusivity $\alpha_{s} = 0.026 \pm 0.00095 \, \mbox{cm}^{2}\mbox{/sec}$. The density measured in this work is $\rho = 1086.54 \, \mbox{kg/m}^3$. The thermal conductivity was obtained from the relation (\ref{relation}) to be $k=2.132 \pm 0.16 \, \mbox{W/mK}$.

Concerning the electric properties of candelilla wax, the time constant for the dark decay of the surface potential found in this work, is an evidence that we are dealing with a very high resistivity material. As we can see from Fig.\ 6, the dark decay of the surface potential obeys quite well the behaviour of an insulator, considered from the point of view of a parallel plate capacitor as presented on section 3.3. In this work we found from the fitting of the data to equation (\ref{sp}) that the time constant $\tau_e=166.17 \pm 5.48 \, \mbox{hours}$, means a resistivity \mbox{$\rho_e=5.98 \pm 0.19 \times 10^{17} \, \mbox{ohm-cm}$} given by equation (\ref{density}).

We can say also that for high resistivity materials, the combination of photoacoustic spectroscopy, the thermal decay method, and the dark decay of the surface potential is a well recommended one, since one can obtain some physical properties of this kind of materials in a very simple way.\\

\vspace{1cm}

{\bf Acknowledgements.} The authors thank Dr.\ J.\ L.\ Mart\'{\i}nez for kindy providing the candelilla wax. This work was partially supported by CONACyT.

\end{document}